\documentclass[aip,jap,preprint]{revtex4-1}
\usepackage{graphicx}% Include figure files
\usepackage{dcolumn}% Align table columns on decimal point
\usepackage{bm}% bold math
\usepackage{wasysym}
\usepackage{psfrag}
\usepackage{setspace}
\input epsf
\newcommand{\figurewidth}{84mm}
\begin{document}
\title{Energetics and diffusion of gold in bismuth telluride}

\author{M.C. Shaughnessy}
\noaffiliation
\author{ J.D. Sugar}
\noaffiliation
\author{ N.C. Bartelt}
\noaffiliation
\author{J.A. Zimmerman}
\noaffiliation

\affiliation{Sandia National Laboratories, Livermore, CA 94551}

\affiliation{Sandia National Laboratories, Livermore, CA 94551}

\pacs{03.67.Mn, 11.25.Hf}
\keywords{thermoelectricty, bismuth telluride, electrical contacts, diffusion, gold}

\date{\today}
\begin{abstract}
We have coupled electron microscopy and energy dispersive spectroscopy experiments  with \textit{ab-initio} modeling to study the solubility and diffusion of Au in Bi$_2$Te$_3$. 
We found that thermal annealing of Au films results in Au concentrations in Bi$_2$Te$_3$ above the previously reported solubility limit.  The time scale of Au diffusion into Bi$_2$Te$_3$ is also much greater than expected.  To explain our observations, we calculate defect formation energies and diffusion barriers within DFT. We identify an interstitial mechanism consistent with the previously observed low solubility and (rapid) anisotropic diffusion.   However, the lower formation energies of substitutional defects suggest  that they may be active in our experiments and  explain the high observed concentrations.

\end{abstract}	

%\pacs{03.67.Mn, 11.25.Hf}

\maketitle

\section{Introduction}
Bismuth telluride, Bi$_2$Te$_3$, is a narrow bandgap semiconductor commonly used for waste heat harvesting and electrical cooling~\cite{Tritt} due to its large thermoelectric figure of merit, ZT. Recent interest in this material has been driven by the prospect of improving ZT through nanostructuring~\cite{Poudel, PbTe_nano, nano_chalcogenides}, as well as by the discovery of topologically protected, highly mobile electronic surface states in Bi$_2$Te$_3$ and related compounds, such as Bi$_2$Se$_3$ and and Bi$_x$Sb$_{1-x}$~\cite{SC_Zhang, Culcer}.  These uses of Bi$_2$Te$_3$ require making electrical contact, so understanding the properties of metal contacts to Bi$_2$Te$_3$ will be valuable for technological progress. In this paper we examine the stability of Au contacts on Bi$_2$Te$_3$.         

\begin{figure}
 \includegraphics[angle=0,width=\textwidth,clip=true]{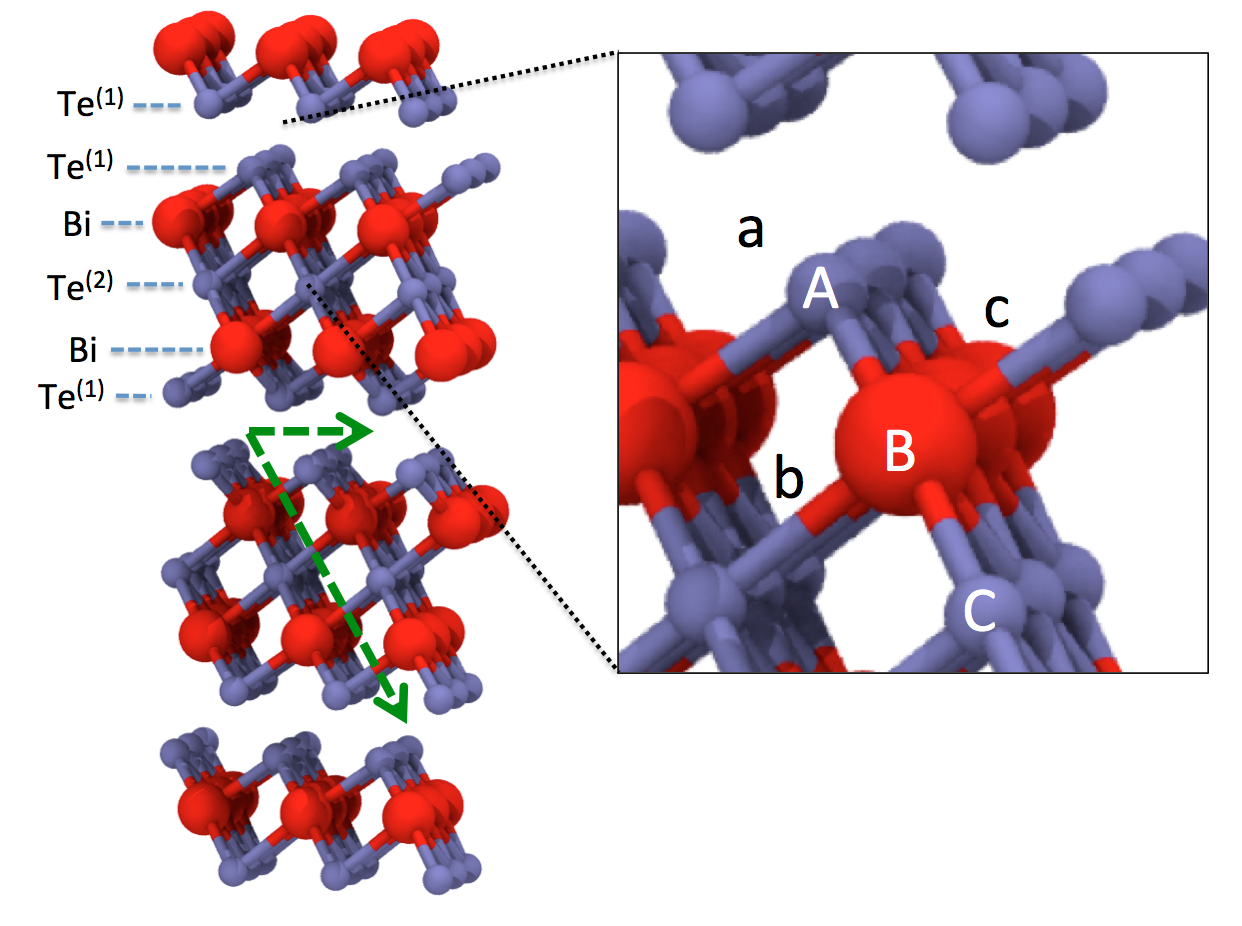}
 \caption{The 135-atom Bi$_2$Te$_3$ unit cell with Te atoms as small purple spheres and Bi atoms as large red spheres. The substitutional doping sites are labels are capitalized: Au$_{{\rm Te}^{(1)}}$ (A), Au$_{\rm Bi}$ (B) and Au$_{{\rm Te}^{(2)}}$ (C). The three possible interstitial sites are labeled in lowercase: Te$^{(1)}$-Te$^{(1)}$ int (a), Bi-Te$^{(2)}$ int (b), Bi-Te$^{(1)}$ int (c). The in-plane (shorter arrow) and out-of-plane (longer arrow) diffusion pathways are indicated by the dashed arrows in the lower portion of the supercell.}
 \label{fig:crystal}
\end{figure}

Various contact metals have been used with Bi$_2$Te$_3$ devices, including Pt~\cite{PtContacts}, Ag~\cite{AgContacts}, and Au~\cite{electrodep_BiTe_on_Au, AuWeso}. Au and Au/Ti layers~\cite{Sandia_barrier, Lin, Bass} in particular have been found to form good electrical contacts with acceptable mechanical properties. However, observations of relatively fast diffusion from Au electrodes into the Bi$_2$Te$_3$~\cite{AuWeso} raises concerns about the use of Au. Bi$_2$Te$_3$ possesses a layered, trigonal structure~\cite{Jenkins, Wiese} in which quintuple layers of Te$^{(1)}$-Bi-Te$^{(2)}$-Bi-Te$^{(1)}$ are stacked in the $z$-direction, as shown in Fig. \ref{fig:crystal}. Measurements of diffusivity and solubility of Au in Bi$_2$Te$_3$ were carried out by Keys and Dutton~\cite{Keys1}. They found fast diffusion parallel to the Te-Te double double layer planes. 
The Au solubility they report is low, about 10$^{18}$ atoms/cm$^3$ at 300K, much lower than that of commonly used dopants in thermoelectric applications~\cite{Yamashita}. The combination of low solubility and fast diffusion might suggest that Au would quickly reach it's solubility limit in Bi$_2$Te$_3$ and that afterwords the Au contact would be stable.%	  

To test this, we monitored the structure of annealed micron-thick Au films on Bi$_2$Te$_3$. We observe fluxes large enough to completely deplete the  Au films. This result is inconsistent with the Keys and Dutton report of low Au solubility. To explain the result we use density functional theory (DFT) to compute the formation energies of isolated Au defects and their diffusivities.
We find interstitial site formation energies and diffusivities are consistent with Keys and Dutton measurements.
However, our calculations also suggest a second slower stage of diffusion, associated with lower energy Au substitution in the Bi$_2$Te$_3$ lattice, that occurs after the initial rapid diffusion, accounting for our experimental results.

This result suggests the solubilities of metals in highly doped, polycrystalline, or vacancy-rich Bi$_2$Te$_3$ may be significantly higher than measured in pristine single-crystal specimens where it is kinetically difficult to form substitutional defects. The concentration of vacancies, which depends on synthesis and processing conditions, controls the solubility of metal dopants. From a practical perspective, Bi$_2$Te$_3$ devices with metal contacts should have a diffusion barrier unless the true equilibrium solubility of the contact metal in the specific Bi$_2$Te$_3$ material being used is known and is acceptable for long-term device performance. 
		           	            
 \section{Experiment}
           
 \subsection{Method}
           
\begin{figure}\label{Pads}
 \includegraphics[angle=0,width=\textwidth,clip=true]{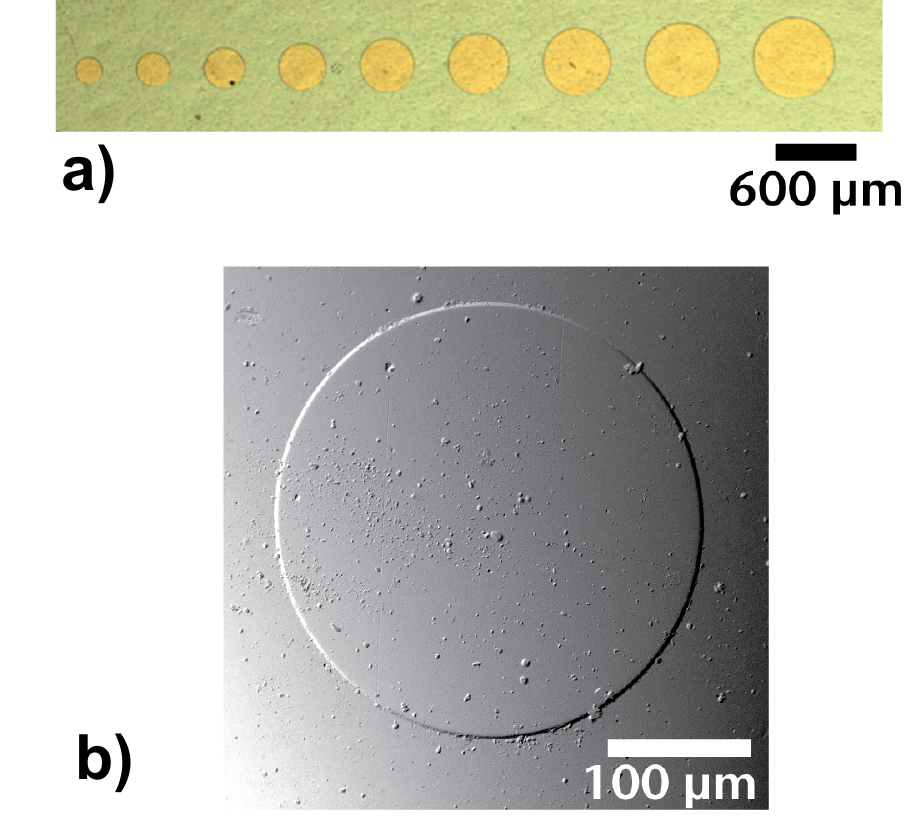}
\caption{The optical micrograph in a) shows a lithographically patterned Au film on n-type Bi$_2$Te$_3$. The Au circles range in diameter from 200 $\mu$m to 600 $\mu$m in steps of 50 $\mu$m.  The SEM micrograph in b) shows a single Au contact as processed. There is cutting debris on the surface because we dice a larger substrate into smaller sections.}
  \label{fig:JDS_pads}
\end{figure}

To perform aging experiments, Au contacts are sputter deposited onto n-type Bi$_2$Te$_3$ substrates. The substrates are sintered and lightly doped with Se to a nominal composition of Bi$_2$Te$_{2.5}$Se$_{0.5}$, improving the thermoelectric properties when compared to undoped Bi$_2$Te$_3$. Circular Au contacts are patterned onto the substrates. Conventional UV lithography is used to pattern circles ranging in diameter from 200 to 600 $\mu$m in steps of 50 $\mu$m onto the substrate surfaces. Before depositing Au, the substrate surfaces are briefly etched with an argon plasma, and then Au is sputter deposited to a thickness of 1 $\mu$m as shown in Fig. \ref{fig:JDS_pads}. Aging is performed in evacuated quartz tubes at 150$^\circ$C, 250$^\circ$C, and 350$^\circ$C for 65 hr and 163 hr. The Au contacts are characterized with scanning electron microscopy (JEOL 7600, JEOL 840, and PhenomWorld desktop SEM) and scanning transmission electron microscopy (JEOL 2010 FEG operating at 200 kV). EDS analysis in the TEM is performed with Digiscan (Gatan, Inc.) and a Si(Li) EDS detector (Oxford, Inc.).
	
 \begin{figure}
 \includegraphics[angle=0,width=\textwidth,clip=true]{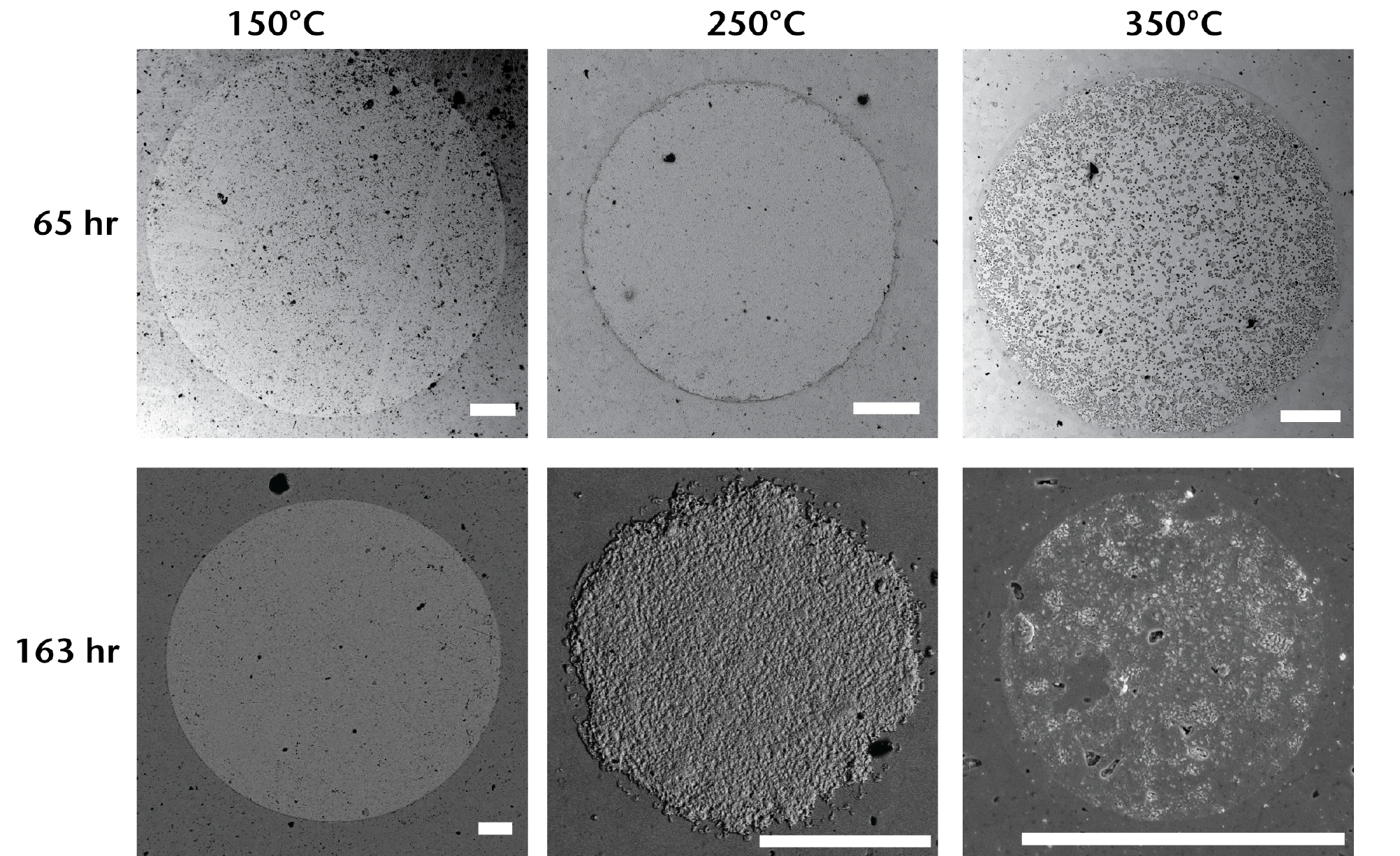}
 \caption{The figure shows a series of Au contacts after various aging conditions (time and temperature). The scale bar in each image is 50 $\mu$m long.} 
 \label{fig:JDS_diff}
\end{figure}     
         
\subsection{Results}
         
 We observe large fluxes of Au into the Bi$_2$Te$_3$ substrate, sufficient to deplete the contact pads (see Fig. \ref{fig:JDS_diff}) after 163 hours at 350$^\circ$C. The degradation is temperature dependent. The Au contacts are intact at 150$^\circ$C up to 163 hours. At 250$^\circ$C there is evidence of degradation around the edges of the sample after 65 hours and roughening and degradation throughout by 163 hours. At 350$^\circ$C the mostly gone by 63 hours and is practically gone after 163 hours. Small islands remain on the surface, and EDS suggests these islands are not pure Au, but contain Bi and Te also. At intermediate aging conditions the edges of the contact seem to disappear first. The contact develops a porous appearance with holes in the middle of the contact. This is most easily visible in the image after 65 hr at 350$^\circ$C.
         
We next show the concentration of Au in Bi$_2$Te$_3$ is several times larger than the previously reported solubility limit~\cite{Keys1}. The concentration of Au in the substrate is computed by dividing the total volume of gold by the substrate volume. We fabricate two samples each with many contact pads, for total Au volumes of $4.2 \times 10^{-5}$cm$^3$ and $3.6 \times 10^{-5}$cm$^3$, corresponding to $2.5 \times 10^{18}$ and $5.1 \times 10^{18}$ Au atoms. The substrates are both disks 1.6 cm in diameter and $L = 0.2$ cm thick. Thus, the Au concentration in the substrate after the Au contact pads are entirely depleted is $6.1 \times 10^{18}$ atoms/cm$^3$ and $ 5.2 \times 10^{18}$ atoms/cm$^3$. The greater value sets a lower limit on the solubility of Au in n-type Bi$_2$Te$_3$ higher than $2 \times 10^{18}$ atoms/cm$^3$ measured in Ref. [\onlinecite{Keys1}]. Further, the time scale for changes in the contact is longer than expected from Ref. [\onlinecite{Keys1}]: Keys and Dutton give $ D \sim 3 \times 10^{-4}$ cm$^2$/sec at 350$^\circ$C, from which one obtains an equilibration time $L^2/D$ of $\sim$ 100 sec, shorter than the hundred hour time scale of the observed changes.
         
 \begin{figure}\label{JDS3}
 \includegraphics[angle=0,width=\textwidth,clip=true]{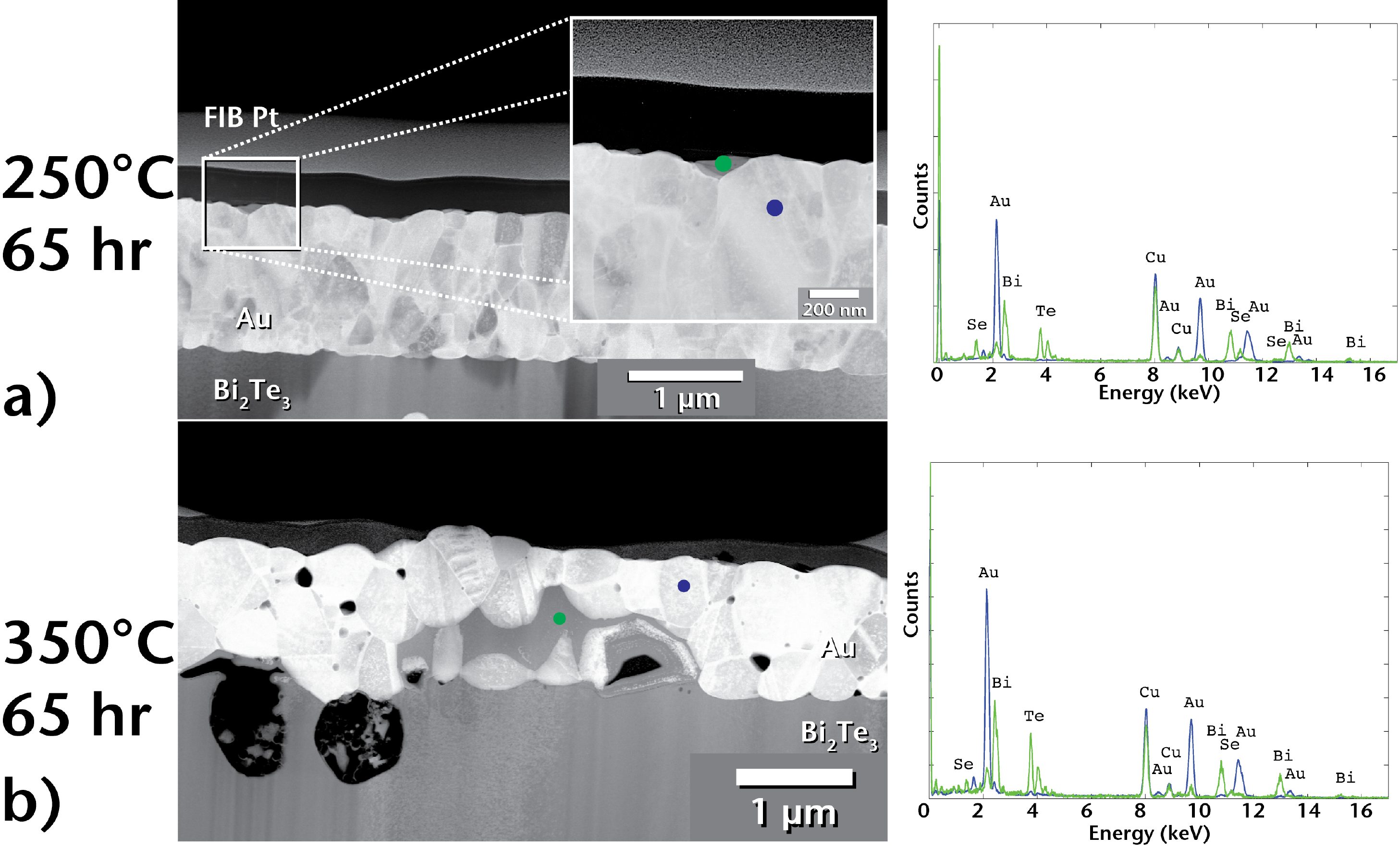}
 \caption{
In a) a HAADF STEM image shows a cross-sectional view of Au on Bi$_2$Te$_3$. Two layers of Pt are visible atop the 1 $\mu$m thick polycrystalline Au contact pad. The inset image is a higher magnification view of the boxed region. In it, there is a feature with different contrast on the surface of the Au at a grain boundary junction. EDS analysis (at the right) on this region and the neighboring Au shows that this feature is a second-phase precipitate crust containing Bi, Te, Se, and Au. The green curve is associated with Bi$_2$Te$_3$ and the blue with Au. It is likely atoms from the substrate diffused along the Au grain boundaries and formed a precipitate at the surface. In b), a second phase precipitate is seen in the middle of the Au film. Again, EDS analysis shows this precipitate contains Au, Bi, Te, and Se. There are also large pores (black) in the substrate and contact.}
\label{fig:JDS_data}
\end{figure}
   
Evidence that non-interstitial sites play a role in the Au diffusion comes from STEM and EDS analysis of the contact/substrate interface shown in Fig. \ref{fig:JDS_data}. Cavities form in both the Au and Bi$_2$Te$_3$, suggesting both substrate and contact atoms migrate. The SEM images in Fig. \ref{fig:JDS_diff}. show pits (dark regions) in the Bi$_2$Te$_3$ substrate under the contact at the later time and highest temperature. EDS analysis of precipitates $\textit{on top}$ of the contact (curves on the right of Fig. \ref{fig:JDS_data}) shows both Te and Bi migrate through the contact and precipitate. Fig. \ref{fig:JDS_data}a shows a precipitate containing both Bi and Te at a grain boundary in the Au at the top surface of the Au.  That Bi and Te are removed form the Bi$_2$Te$_3$ lattice suggests the possibility of substitutional Au incorporation into Bi$_2$Te$_3$.  The energetics of this scenario is investigated in  the calculations of substitutional formation energies presented below. 

\section{Modeling}
           
The goal of this section is to understand why the Keys and Dutton \cite{Keys1} measurements fail to explain our experiments.  To do this we perform DFT calculations of Au diffusion in Bi$_2$Te$_3$. We first show that interstitial doping is consistent with Keys and Dutton.  We then show that Au substitutional sites are likely to have lower formation energy.
           
\subsection{Method}
           
We use a 135-atom supercell of  Bi$_2$Te$_3$ for density functional theory (DFT) calculations. Previous studies of native defects in Bi$_2$Te$_3$ suggest such a supercell can accurately model isolated defects \cite{Antisite}. The VASP code \cite{VASP}, with standard projector augmented wave (PAW) psuedopotentials, with d-electrons in the core, is used to compute the electronic structure. A planewave energy cutoff of 500 eV and a Monkhorst-Pack \textbf{k}-point mesh of $2 \times 2 \times 2$ are adopted. The basis vectors and ionic positions are relaxed to determine the low energy crystal structure, whose a = 4.35 \AA\ and c = 29.84 \AA\ lattice constants, as well as the Te-Te layer separation (3.46 \AA) are within 1\% of the experimental values ($a$ = 4.38 \AA\ and $c$ = 30.36 \AA\ Te-Te distance = 3.63 \AA\ \cite{BiTe_exp_latticevectors, Wiese}). Relaxations are finished when all forces on all atoms are less than 6 meV/\AA.   In the subsequent doping studies, the dimensions of the supercell are held fixed and only ionic relaxations are allowed. The spin-orbit interaction is taken into account because the Bi and Te have heavy nuclei. Figure \ref{fig:crystal} shows the pure Bi$_2$Te$_3$ crystal supercell. There are three unique crystal atom sites and three interstitial sites, although sites a and c turn out to be energetically equivalent for Au doping.
                  
We consider two types of substitutional defect. In one type a Au atom replaces a single Bi or Te atom and the replaced atom is removed entirely. In the second type a Au atom occupies a Bi or Te site and pushes the Bi or Te atom into an adjacent interstitial site. Interstitial defects are also modeled by adding an Au atom into the pure bismuth telluride supercell at one of the three possible interstitial sites with no initial displacement of any atoms.   

The interstitial defect formation energy is defined \cite{Northrup} as:
\begin{equation}
\Delta E_f = E_{doped} - (\mu_{Au} + E_{pure})
\end{equation}
where E$_{doped}$ is the total energy of the supercell with the dopant, $\mu_{Au} $ is the chemical potential for Au, computed from fcc bulk Au, and E$_{pure}$ is the total energy of the Bi$_2$Te$_3$ supercell with no dopant. $\Delta$E$_f$ is the formation energy, the energy penalty to create the doped system from the pure crystals. 

In the case of a substitutional defect, we assume the displaced host crystal atom (Bi or Te) is sent to a chemically well-defined reservoir. Then,
 \begin{equation}
\Delta E_f = E_{doped} - (\mu_{Au} + E_{pure}-\mu_{Te/Bi})
\end{equation}
where $\mu_{Te/Bi}$ is the chemical potential of the displaced Bi or Te atom in the reservoir. If the displaced atom remains at the adjacent interstitial site we revert to Eq. 1.
 
In a typical Bi$_2$Te$_3$ application, such as in the active regions of a power generator or cooler, there could be a variety of chemical environments and hence chemical potentials. Because the Au-Bi-Te phase diagram is known \cite{ciobanu, Winkler}, we assume the system of Bi$_2$Te$_3$ with the Au dopant is in equilibrium with pure Au (from a contact, say) and either AuTe$_2$ or BiTe. Depending on composition, Bi$_2$Te$_3$ can coexist with, AuTe$_2$, BiTe, or Au \cite{Kobachevski}.  

AuTe$_2$ crystalizes in the calaverite structure with a three atom unit cell~\cite{Calaverite}. The BiTe~\cite{BiTe} is one of many possible bismuth-rich structures \cite{Bos2}; it is related to the tetradymite Bi$_2$Te$_3$ structure, with an additional Bi double layer inserted between every other Te$^{(1)}$-Te$^{(1)}$ double layer.
 
The three $\mu_{Au}$, $\mu_{Bi}$, $\mu_{Te}$ can be determined from the total energies of Au, AuTe$_2$, BiTe, and Bi$_2$Te$_3$. We solve the first along with either the second or the third of the equations below.
\begin{equation}
2\mu_{Bi}+3\mu_{Te} = \mu_{Bi_2Te_3}
\end{equation} 
\begin{equation}
\mu_{Bi}+\mu_{Te} = \mu_{BiTe}
\end{equation}
\begin{equation}
\mu_{Au} + 2\mu_{Te} = \mu_{AuTe_2}
\end{equation}
where $\mu_{Au}$, $\mu_{AuTe_2}$ $\mu_{BiTe}$ and $\mu_{Bi_2Te_3}$ are the total energies of one formula unit of Au, AuTe$_2$, BiTe, or Bi$_2$Te$_3$, respectively. %In the Bi-rich case, Eq. 3 and Eq. 4 are solved, while in the Te-rich case Eq. 3 and Eq. 5 are solved. 

Since there is no ternary phase, nor a relevant BiAu phase, our choice of chemical reservoirs is nearly unique for the dilutely Au doped stoichiometric Bi$_2$Te$_3$. (We note above $116\, ^{\circ}\mathrm{C}$, BiAu$_2$ is entropically stabilized \cite{Chevalier} and so it may appear as a possible reservoir. Errors associated with neglecting this phase would be on the order of thermal energies, i.e., small.) 

The activation energy for diffusion of Au in the in-plane direction between adjacent Te-Te double layer interstitial sites is calculated with the nudged elastic band method (NEB)~\cite{Mills_NEB}. Au can diffuse by other ways in bismuth telluride. Along with diffusion in the cross-plane or z-direction, more complicated mechanisms, such as vacancy mediated diffusion, multiple-defect diffusion or even diffusion along grain boundaries in polycrystalline samples may contribute to the experimentally measured diffusion. The calculations presented here provide an upper bound on the diffusion through single crystal Bi$_2$Te$_3$ along solely interstitial paths. 

The starting point for the NEB calculations is an Au atom in the Te-Te double layer interstitial site in the 135-atom supercell. The final configuration has the Au atom moved to an adjacent Te-Te double layer interstitial site, either in the same plane or in an adjacent Te-Te double layer. Initially, a series of 16 equally-spaced atomic positions for the Au along a line connecting the initial and final Au positions are the assumed diffusion pathway. Then, forces on each of the 16 images of the isolated Au atom along the line are computed. All atoms in each of the 16 supercells relax, with the caveat that an additional fictitious force acts on each Au atom. This elastic band force is due to adjacent images of the Au atoms, keeping them separated. In this way, the Au image atoms are forbidden from simply relaxing to the globally favored initial or final configuration sites. The forces on all the atoms, including the fictitious elastic band forces, in each of the image supercells are recomputed until self-consistent cycle is complete. In this way low-energy metastable saddle point-crossing pathways can be found in the vicinity of the initial guess for the diffusion pathway. 
           
\subsection{Results}
         
\begin{figure}\label{pathway}
 \includegraphics[angle=0,width=\figurewidth,clip=true]{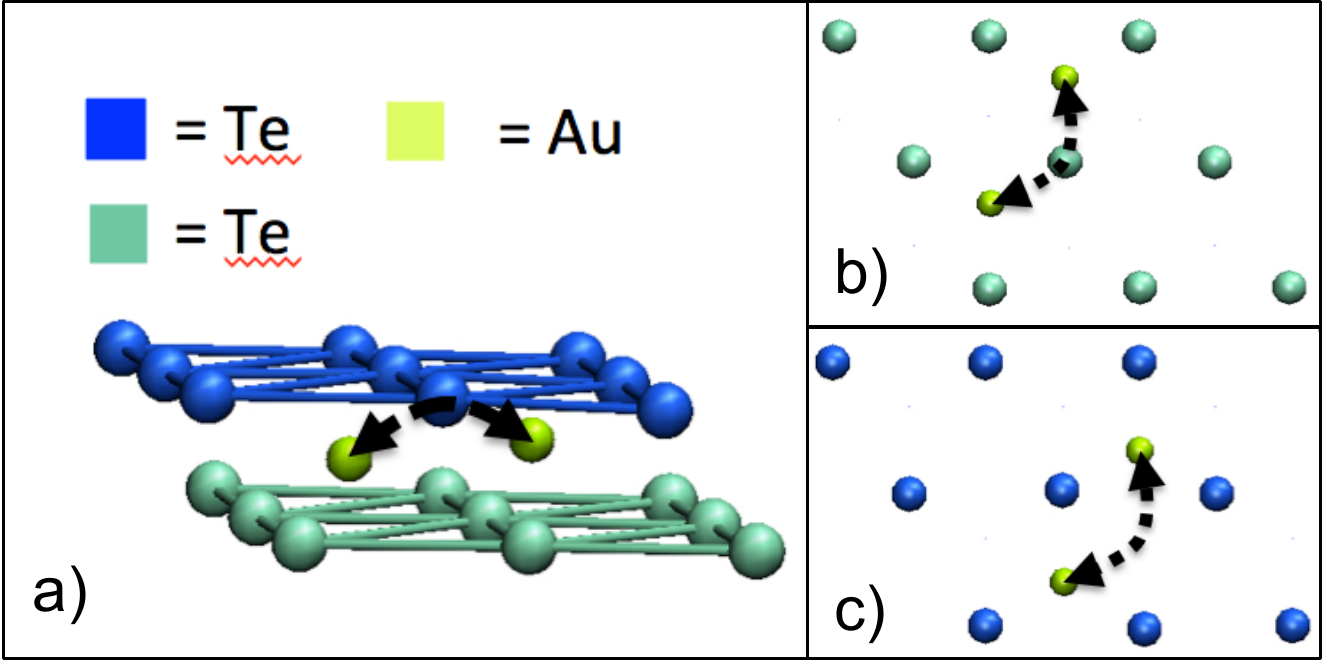}
 \caption{Schematic illustrating the Au in-plane diffusion pathway in the Te-Te double layer. In a) only the Au atom (lime sphere) and the Te atoms in the Te-Te double layer are shown for clarity. In b) and c) the view is down the c-axis and only the Te atoms in the lower or upper Te layer, respectively, are shown.  The Au atom is shown at the initial and final positions. The pathway bends upward out of the plane at its midpoint to avoid the Te atom in the lower plane.}
 \label{fig:pathway}
\end{figure}

\begin{figure}\label{NEBplot}
  \includegraphics[angle=0,width=\figurewidth,clip=true]{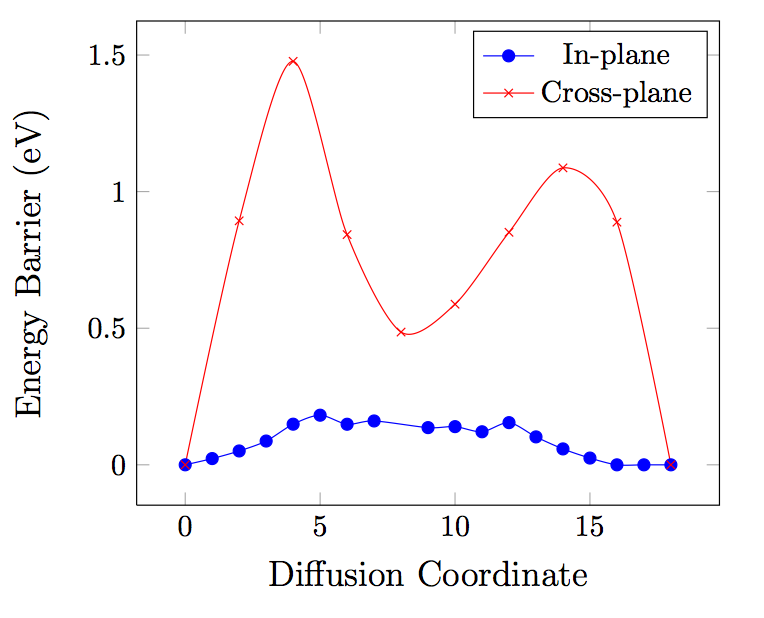}
  \caption{Total energy as a function of position along the diffusion pathway. The initial Te-Te interstitial site corresponds to reaction coordinate 0. The energies are not symmetric about the midpoint because of the underlying asymmetry in the crystal.}
  \label{fig:NEBplot}
\end{figure}
		
Considering first interstitial diffusion in the Te-Te double layer plane, we compute the in-plane diffusion barrier and find it is 0.18 eV, in good agreement with the experimental value of 0.20 eV \cite{Keys1}. The diffusion is not along a straight line connecting the adjacent interstitial sites. In the relatively open structure of the Te-Te double layer, the Au atom moves along a curved path to maximize the Te-Au distance. Fig. \ref{fig:pathway} shows the pathway, viewed along the c-axis with only the equilibrium positions of the Te atoms in the relevant double layer shown. The Te atoms slightly distort to allow the Au atom to squeeze past, with the largest distortions about one-quarter and about three-quarters of the way along the transit. Fig. \ref{fig:NEBplot} shows the total energy change along the diffusion pathway. 
The good agreement between experimental and theoretical values of the in-plane activation energy and diffusivity (in the next section) suggest the in-plane diffusion in highly crystalline samples is indeed through the interstitial mechanism discussed here. 

We also compute the activation energy for diffusion in the out-of-plane direction to reach an adjacent Te-Te double layer. In this case the barrier is nearly 1.5 eV, occurring where the Au atom leaves the Te-Te double layer and pushes into the interior of the quintuple layer. The large barrier and the associated disagreement with the experimentally measured value of 0.56 eV~\cite{Keys1} suggest other mechanisms must allow the Au atom to move in the out-of-plane direction more easily. Given the low formation energies for substitutional defects  diffusion pathways involving substitutional Au intermediate states could be involved, either through vacancies or Bi- and Te- antisite defects. If such pathways are relevant they would be expected to lower the diffusion barrier because the substitutional defect would decrease the strain in the lattice. That the calculated energy barrier in the cross plane direction is largest ($\sim$1.5 eV) as the Au is leaving the Te-Te double layer suggests other interstitial-type paths between Te-Te double layers will encounter a similar barrier, so some pathway not accessible to the interstitial NEB method seems to be required. 

In addition to the barrier height, the diffusivity itself $D = D_0 e^{E_{act}/k_BT}$ for the Au diffusing through the Bi$_2$Te$_3$ along the Te-Te double layer can be computed.using $D_0 = \lambda a_0^2 \nu e^{\Delta S/k_B}$ ~\cite{Zener}. Here $\lambda$ is a geometrical factor taking into account the various paths the diffusing particle can take, $a_0$ is the lattice constant and $\nu$ is the vibrational frequency of the diffusing particle in the crystal potential field. In a harmonic approximation $\nu$ is approximately equal to $\sqrt {E_{act}/2ma^2}$ where $E_{act}$ is the activation energy (barrier height), $m$ is the Au atom mass and $a$ is the jump distance. $\Delta S$ is an entropy term to account for the changing strain in the lattice due to the diffusing particle. It is reasonable to assume $\Delta S$ is zero, given the open Bi$_2$Te$_3$ structure and the small distortion of the lattice \cite{Zener, Cu_BiTe_diffusion}. For forward diffusion in the Bi$_2$Te$_3$ Te-Te double layer, $\lambda = 3/2$; using $a_0$ = 4.38 \AA\ and $\nu$ = 0.49$\times$10$^{12}$ s$^{-1}$. Then the calculated $D_0$ = 1.41$\times$10$^{-3}$ e$^{\Delta S/k_B}$ cm$^2$/s. The experimental value is 1.13$_{-0.31}^{+0.54}\times$10$^{-3}$ cm$^2$/s \cite{Keys1}, in good agreement with the calculated value for the in-plane interstitial mechanism, suggesting this mechanism dominates transport parallel to the Te-Te double layer directions (and also that $\Delta S$ is indeed nearly zero). 
         
Table 1 gives defect formation energies for interstitial, substitutional, and substitutional with displaced Bi or Te atoms. The formation energies with respect to the two reservoirs are approximately the same. The lowest interstitial formation energy is the Te-Te double layer site, Te$^{(1)}$-Te$^{(1)}$ int,  because of the relatively open structure and weak bonding between the Te-Te double layers compared to the other possible sites. Another interstitial site, the Bi-Te$^{(1)}$ int site, is unstable for Au doping. At this site the Au atom relaxes into the region between Te-Te double layer to the Te$^{(1)}$-Te$^{(1)}$ int site. The Bi-Te$^{(2)}$ interstitial site formation energy is higher because the Au atom must distort more of the rigid covalent bonds present in the quintuple layer. 
         
The simple substitutional formation energies are lower than either interstitial defect formation energy, indicating these should be more prevalent.   The calculated values depend only weakly on the Bi and Te chemical potentials; for all conditions the substitutional formation energies are relatively low. Exceptionally low formation energy for substitutional defects might be expected     based on recent calculations on the phonon bandstructure of AuBiTe$_2$ \cite{Vidvuds}.  The host compound is hardly stable and substitution should be energetically cheap.      Experiments~\cite{Taylor_vapor} also show that the composition of near-stoichiometric solid Bi$_2$Te$_3$ material can be easily varied by changing the Te vapor pressure due to a low Te sublimation energy. This low Te sublimation energy poses a problem when Bi$_2$Te$_3$ is annealed at high temperatures and the Te sublimes. Both these modeling and experiment suggest substitutional defects should be easy to form.
      
Considering the formation energies for the substitutional models with an internally displaced Te atom, their large magnitude can be understood because the displaced Te atom relaxes into the Te-Te double layer disrupting the van der Waals bonding between the Te-Te layers. The configuration with the Au substituting for a Bi and displacing the Bi into an adjacent interstitial site does not exist because it is unstable; the Bi atom returns to its crystal site and the Au relaxes into the interstitial region. The large formation energies of these substitutional-interstitial defect complexes suggests the occupation of the substitutional sites may be kinetically limited. In other words, the formation of an interstitial defect will be necessary in order to have an Au atom at a substitutional site, so even though substitution is energetically favored overall, it may take a very long time for the Au to reach the substitutional sites because of the large energy barrier. This kinetic mechanism also favors the interstitial pathway for diffusion, especially in highly crystalline samples, because while the formation energy for interstitials is large compared to substitutional defects, the effective formation energy for substitutional sites will be larger. 
\begin{table}
\label{tab:1}
\begin{center}
\begin{tabular}{ccc}
\hline
\hline
&Site &Formation Energy (eV)\\
\hline
&Te$^{(1)}$-Te$^{(1)}$ int & 1.0 eV  \\
&Bi-Te$^{(2)}$  int & 1.13 eV    \\
\hline
AuTe$_2$ Reservoir & Au$_{\rm Bi}$ & 0.53 \\
&Au$_{{\rm Te}^{(1)}}$  &  0.58\\
&Au$_{{\rm Te}^{(2)}}$ &  0.41\\
\hline
BiTe reservoir & Au$_{\rm Bi}$ & 0.48 \\
&Au$_{{\rm Te}^{(1)}}$ &  0.62 \\
&Au$_{{\rm Te}^{(2)}}$ &  0.45 \\
\hline
%&Au$_{Bi}$-Bi int & 1.13  \\
& Au$_{{\rm Te}^{(1)}}$-Te int  & 3.00 \\
& Au$_{{\rm Te}^{(2)}}$-Te int & 3.42 \\
\hline
\hline
\end{tabular}
\end{center}
\caption{Single Au-dopant formation energies. Substitutional formation energies are computed either by displacing substituted lattice atom into an adjacent interstitial site (or Au$_{Te^{(X)}}$) or by using $\mu_{Bi}$ and $\mu_{Te}$ derived assuming the system contacts reservoirs of Au and either AuTe$_2$ or BiTe.}
\end{table}

The calculations yield the formation energies for Au dopants in the perfectly crystalline Bi$_2$Te$_3$ environment, but Au-dopants will coexist alongside native defects in real Bi$_2$Te$_3$. Antisite defects are expected to be dominant, but vacancies and self-interstitials may be present \cite{Antisite, Horak}. These various native defects will allow for different Au doping mechanisms because of the relatively small Au interstitial and substitutional defect formation energies. The processing involving deformation and intentional p- or n-type doping~\cite{Cu_doping} could further alter the population and solubility of Au dopants. Diverse Cu doping locations in Bi$_2$Se$_3$, a closely related material with an identical crystal structure and a van der Waals bound Se-Se double layer~\cite{Easy_Cu_BiSe2}, demonstrate a sensitive dependence on processing conditions. Because the diffusion of Au from a contact is inherently non-equilibrium, the diffusing Au atoms in any case should not be expect to exclusively occupy the lowest energy substitutional sites.    
   
In spite of the potential for many coexisting Au doping sites, the weak bonding and large spacing between the Te-Te double layers suggest Au diffusion through Te-Te interstitial sites in Bi$_2$Te$_3$ could be important over a range of doping and processing conditions. Experimental measurements of diffusion barriers of Au \cite{Keys1}, Ag \cite{Keys2} and Cu \cite{Cu_BiTe_diffusion} at room temperature in single crystal Bi$_2$Te$_3$ indicate the diffusion is strongly anisotropic, with an activation energy of 0.20 eV (0.45, 0.21 eV) for the in-plane direction and 0.56 eV (0.92, 0.80 eV) in the cross plane direction for Au (Ag,Cu). 

\section{Conclusion}
                     
We observed diffusion of Au into n-type Bi$_2$Te$_3$ substrates and migration of Bi and Te on to the top of the Au contact. We have calculated formation energies, diffusion barriers and diffusivity for a likely path for single Au atoms diffusing in Bi$_2$Te$_3$. The lowest energy interstitial defect is between the Te-Te double layer. The other stable interstitial site is in the quintuple layer between the Bi and Te layer. Substitutional defects have low formation energies for Bi$_2$Te$_3$ in equilibrium with Au and either AuTe$_2$ or BiTe chemical reservoirs.  Au substitutional defects with an adjacent displaced Bi or Te atom are found to have larger formation energies than interstitial defects due to the presence of, essentially, two defects.            
           
Calculated values for the diffusion barrier and diffusivity along the in-plane direction agree well with experiment. The diffusion pathway for motion in the parallel direction winds through the Te atoms in the double layer. For diffusion in the perpendicular direction, the barrier is large and the agreement between the calculated value and experiment is poorer. The discrepancy may be due to diffusion pathways beyond the reach of the NEB method we employ. Experimental measurements of anisotropic diffusion of Cu \cite{Cu_BiTe_diffusion}, Ag \cite{Keys2}, and Se \cite{Se_diffusion_mechanism} in Bi$_2$Te$_3$ suggest antisite defects, thermal vacancies and interstitial mechanisms can be important. 
                     
Our results suggest that in Bi$_2$Te$_3$ near stoichiometry, if and when equilibrium is reached, the majority of Au atoms should occupy substitutional sites and any Au interstitial defects should predominately be found in the Te-Te double layer. The initial diffusion of Au should occur mainly along the in-plane direction by a mechanism involving the interstitial sites between the Te-Te double layer. When a diffusing Au atom occupies a low-energy substitutional site, it will cease moving quickly. Because of the kinetic barrier associated with moving the displaced Te or Bi atom into the interstitial, however, it may be relatively rare for an Au atom to occupy a substitutional site. If an Au atom occupies a Te-Te double layer interstitial site, it may be expected to rapidly diffuse along the in-plane direction. Our experiment on Au in contact with Bi$_2$Te$_3$ demonstrates the rapid diffusion of Au continuing beyond the solubility limit suggested in Ref. [\onlinecite{Keys1}]. We suggest the diffusion occurs in two stages. First, Au rapidly diffuses along the Te-Te double layers then quickly ceases to diffuse as the low solubility limit for interstitial Au is reached. Next, slower diffusion, through either a vacancy- or interstitial-mediated process occurs, filling the substitutional sites. The timing of the onset of the second stage might be expected to depend sensitively on the initial composition of the Bi$_2$Te$_3$; for high quality single crystal specimens it would be exceedingly slow because of a lack of vacancies and other defects to aid the formation of substitutional Au defects, while for more highly defective polycrystalline materials the second stage would occur more quickly. In either case, because we observe migration of Bi and Te out of the substrate, we expect new vacancies to be formed over time and allow for continued diffusion of Au into the substrate, beyond the solubility limit of Ref. [\onlinecite{Keys1}].  This incorporation of Au into the active thermoelectric material may degrade carefully tuned doping and transport properties. Further, the motion of Te and Bi into and onto the Au film creates voids at the interface, substantially reducing the effective contact area.

%While for highly crystalline materials the second stage substitutional diffusion is expected to be kinetically slowed by the large cost of forming host interstitial - Au substitutional defect complexes, these materials might also be expected to show larger deviations in thermoelectric performance due to incorporation of Au than would more polycrystalline or off-stoichiometric samples. 
            
\section{Appendix}
 Because thermoelectric devices are usually exposed to significant temperature gradients, it is useful to consider diffusion under such conditions. A temperature gradient introduces an additional term in the expression for diffusive flux \textit{\textbf{J}} in a concentration field, \textit{c}. 
\begin{equation}
\mathbf{J} = -D\frac{\partial c}{\partial \mathbf{x}} - Dc\frac{Q^*}{k_BT^2}\frac{\partial{T}}{\partial{\mathbf{x}}}.
\end{equation}
where $Q^*$ is the heat of transport. The second term on the right captures the Soret effect. We estimate $Q^*$ as follows. The simple interstitial transport mechanism has a relatively low activation barrier and so we assume this mechanism dominates transport in the double layer. Because the $\Delta S$ in the diffusivity is small, and no covalent bonds are formed or broken, we further simplify the estimate of $Q^*$ by assuming the phonon spectrum along the diffusion path is unchanging. In this approximation -E$_{act}$ $\le$ $Q^*$ $\le$ $E_{act}$ \cite{Asaro}; the limits are satisfied when the activation energy is dissipated about the initial site, or when all of it is instead carried to the final site. If the activation energy is evenly dissipated between the initial and final sites then there is no energy flux as the particle moves and $Q^* = 0$. In the present case the diffusion barrier profile, with the largest drop in energy occurring near final site, suggests energy will be mostly dissipated near the final site and so we estimate $Q^*$ $\sim$ 0.18 eV (the calculated diffusion barrier) for in-plane thermomigration. Experimental measurement of 0.265 eV heat of transport for Ag in Bi$_2$Te$_3$\cite{Dibbs} suggests the sign and magnitude of this estimate are reasonable.  
            
\section{Acknowledgements}
Sandia National Laboratories is a multi-program laboratory managed and operated by Sandia Corporation, a wholly owned subsidiary of Lockheed Martin Corporation, for the U.S. Department of Energy's National Nuclear Security Administration under contract DE-AC04-94AL85000.
            
\bibliography{Au_BiTe}          
          
\end{document}